\documentclass[12pt,preprint]{aastex}

\newcommand\psr{PSR~J0737$-$3039}
\def\lapp{\ifmmode\stackrel{<}{_{\sim}}\else$\stackrel{<}{_{\sim}}$\fi}
\def\gapp{\ifmmode\stackrel{>}{_{\sim}}\else$\stackrel{>}{_{\sim}}$\fi}

\begin{document}

\title{Green Bank Telescope Observations of the Eclipse of Pulsar ``A'' in the
Double Pulsar Binary PSR J0737$-$3039}

\author{
V. M. Kaspi,\altaffilmark{1,2,3}
S. M. Ransom,\altaffilmark{1,2}\\
D. C. Backer,\altaffilmark{4}
R. Ramachandran,\altaffilmark{4}
P. Demorest,\altaffilmark{4}
J. Arons,\altaffilmark{4,5}
A. Spitkovsky,\altaffilmark{6,5,7}
}

\altaffiltext{1}{Department of Physics, Rutherford Physics Building,
McGill University, 3600 University Street, Montreal, Quebec,
H3A 2T8, Canada}

\altaffiltext{2}{Department of Physics and Center for Space Research,
Massachusetts Institute of Technology, Cambridge, MA 02139}

\altaffiltext{3}{Canada Research Chair; NSERC Steacie Fellow}

\altaffiltext{4}{Department of Astronomy, University of California, 601 Campbell Hall, Berkeley, CA 94720}

\altaffiltext{5}{Theoretical Astrophysics Center and Physics Department, University of California, 601 Campbell Hall, Berkeley, CA 94720}

\altaffiltext{6}{KIPAC, Stanford University, P.O. Box 20450, MS 29, Stanford, CA 94309}

\altaffiltext{7}{Chandra Fellow}

\begin{abstract}

We report on the first Green Bank Telescope observations at 427,
820 and 1400~MHz of the newly discovered, highly inclined and relativistic
double pulsar binary.  We focus on the brief eclipse
of PSR~J0737$-$3039A, the faster pulsar, when it
passes behind PSR~J0737$-$3039B.  We
measure a frequency-averaged eclipse duration of $26.6 \pm 0.6$~s, or
$0.00301 \pm 0.00008$ in orbital phase.  The eclipse duration
is found to be significantly dependent on radio frequency, with eclipses longer
at lower frequencies.  Specifically, eclipse duration
is well fit by a linear function having slope 
$(-4.52 \pm 0.03) \times 10^{-7}$~orbits~MHz$^{-1}$.
We also detect significant asymmetry in 
the eclipse.  Eclipse ingress takes $3.51\pm 0.99$ times longer than egress,
independent of radio frequency.  Additionally, the eclipse lasts 
$(40 \pm 7) \times 10^{-5}$ in orbital phase
longer after conjunction, also independent of frequency.  We detect significant
emission from the pulsar on short time scales during eclipse in some orbits.
We discuss these results in the context of a model in which
the eclipsing material is a shock-heated plasma layer within the slower
PSR~J0737$-$3039B's light cylinder, where the relativistic pressure of
the faster pulsar's wind confines the magnetosphere of the slower
pulsar.

\end{abstract}


\section{Introduction}

The discovery of the double-pulsar binary system
\psr\ \citep{bdp+03,lbk+04} represents a landmark in neutron-star
astrophysics.  This relativistic system consists of two neutron stars,
both radio pulsars, in a 2.4-hr orbit having eccentricity 0.08, viewed
nearly edge-on.  The binary promises unparalleled tests of General Relativity.
During each orbit, the faster-rotating of the two pulsars,
PSR~J0737$-$3039A (hereafter A), which has period $P=22$~ms, is
eclipsed briefly at conjunction, while the slower-rotating of the two,
PSR~J0737$-$3039B ($P=2.7$~s; hereafter B), has unusual and longer-lived
flux enhancements near the same epoch \citep{lbk+04}.  This bizarre and
unprecedented behavior offers a new opportunity to understand pulsar
magnetospheric and wind structures.

This paper is part of a short series describing the first follow-up
observations of this remarkable source using the 100-m diameter National 
Radio Astronomy Observatory Green Bank
Telescope (GBT).\footnote{ The National Radio Astronomy Observatory is
a facility of the National Science Foundation operated under
cooperative agreement by Associated Universities, Inc.}  Here we
concentrate on the properties of the pulsar ``A'' eclipse.  From
observations at the Parkes and Lovell Telescopes, \citet{lbk+04} have
already established that the eclipse has duration 20--30~s and is
roughly independent of frequency.  The greater sensitivity of the
observations described here reveal that the eclipse duration is in
fact slightly but significantly dependent on radio frequency,
as well as demonstrate for the first time that the eclipse is 
asymmetric.

\section{Observations}

Our observations with the GBT were promptly
scheduled as a NRAO Rapid Response Science
Program\footnote{http://www.vla.nrao.edu/astro/prop/rapid/}
four times in 2003 December and once in 2004
January and made use of receivers at 427, 820 and 1400~MHz.  
The observational details are summarized in Table~\ref{ta:obs}.
The 427 and 820 MHz receivers were located at the prime
focus of the GBT, are cooled FET amplifiers, and have
approximate system temperatures of 57~K and 25~K, respectively.  The 1400
MHz receiver was located at the Gregorian focus and is a cooled HFET
amplifier with system temperature of 20~K.  All provide one beam
and are sensitive to orthogonal polarizations.  Signals from the
receivers are transferred from the telescope to the control room via
optical fibre.

The data reported on here were acquired with a single Berkeley-Caltech
Pulsar Machine (BCPM).  The BCPM is an analog/digital filter bank which
samples $2 \times 96$ channels using 4 bits at flexible sampling rates
and channel bandwidths \citep{bdz+97}.  For our observations, the two
orthogonal polarizations were summed in hardware prior to being written
to disk.  The sampling time for all observations was 72~$\mu$s.
The total bandwidth used (and hence the channel bandwidth) depended on
the observing frequency (Table~\ref{ta:obs}).

\section{Analysis and Results}

The data were analyzed using the {\tt PRESTO} software package
\citep{ran01}.  We dedispersed the data using dispersion measure
48.9115~pc~cm$^{-3}$ and did the following analysis on the resulting
time series.  First, we folded data obtained within 2~min of pulsar A's
eclipse with the ephemeris reported by \citet{lbk+04} using 64 phase
bins.  The resulting profile was cross-correlated with a high
signal-to-noise template (obtained using data from several complete
orbits at the appropriate radio frequency) and the temporal offset was
recorded.  Then, we folded data from within the 2~min of the eclipse
again, but in 2-s intervals.  We then determined the pulsed intensity
in each 2-s interval, first by aligning the pulse with the template
using the previously recorded offset, then by finding the best-fit
scale factor between the template and aligned data profile.  The
best-fit value was found by $\chi^2$ minimization, and the equivalent
$1\sigma$ uncertainties determined from the values of the scale factor
corresponding to the minimum $\chi^2$ plus 1.  Note that no effort was
made to calibrate the intensities in physical units; we assume that the
telescope gain and system temperatures were constant throughout each
observation.  The average pulse arrival time in each 2-s interval was
recorded and reduced to the solar system barycenter using the published
position for the pulsar \citep{lbk+04}.  In this way, we obtained a
pulsed flux intensity time series consisting of a barycentric time and
an intensity for each of the 120 2-s intervals around conjunction.  The
number of conjunctions observed at each frequency is given in
Table~\ref{ta:eclipse}.  Note that we found no evidence for any changes
in dispersion measure near eclipse.  Using the 427~MHz data, we set an
upper limit on any dispersion-measure variation on $\gapp 2$-s time
scales of 0.016~pc~cm$^{-3}$.  This was done by running the pulsar
timing software package {\tt
TEMPO}\footnote{http://www.atnf.csiro.au/research/pulsar/timing/tempo/}
on the 427~MHz data (with eclipse points identified by large arrival-time 
uncertainty and omitted) without fitting for any parameters, and
taking the largest residual as an upper limit on any delay due to an
anomalous dispersion-measure change.

\begin{figure}
\plotone{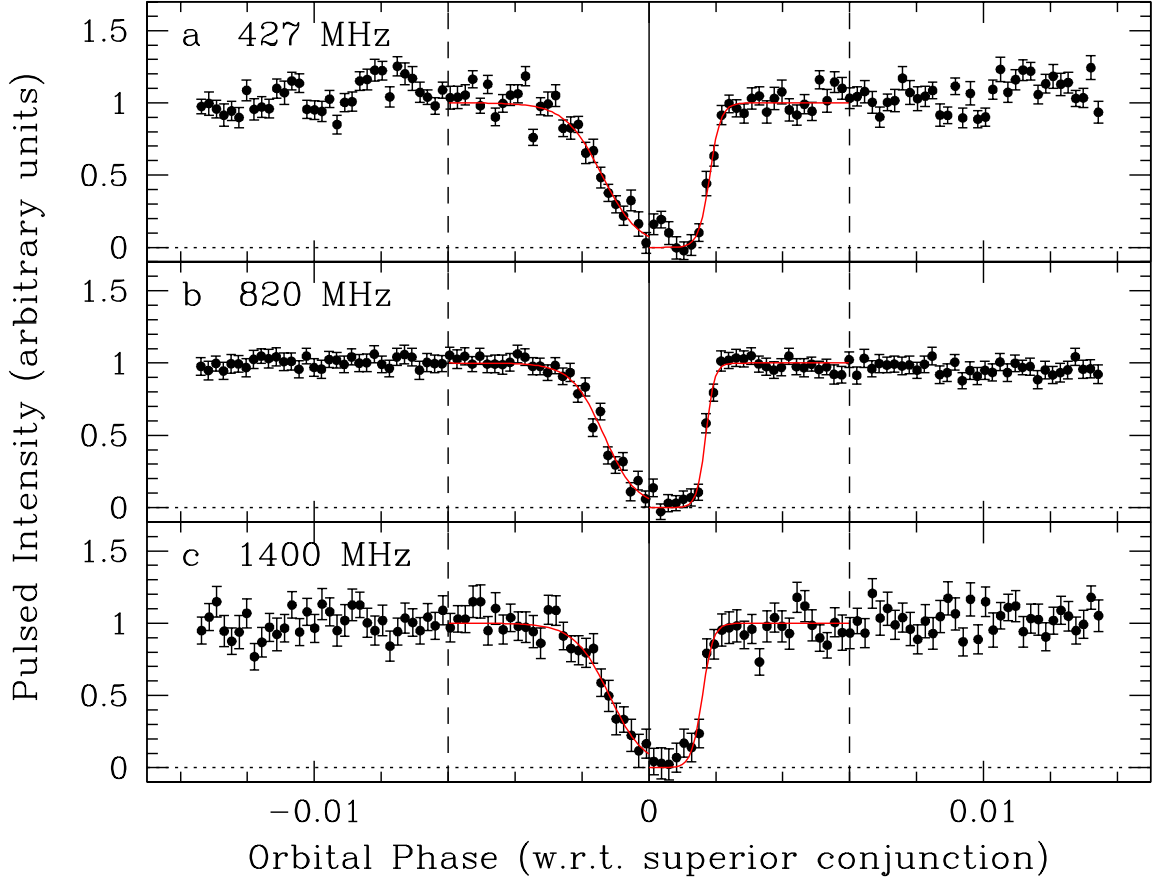}
\figcaption{Pulsar A eclipse light curves.  Each point represents 2~s
of data; the shown curve is 4-min in duration, centered on 
conjunction.  The x-axes are orbital phase with respect to 
conjunction.  The y-axes are pulsed flux, normalized such that the
pre-eclipse flux is unity.  The panels are for (a) 427 MHz, (b) 820
MHz, and (c) 1400 MHz.
The solid vertical line indicates conjunction
and the horizontal dotted lines show 0 flux.  Vertical dashed lines at
$\phi=-0.006, +0.006$ indicate the range of data fitted.  The best-fit
ingress and egress model curves are shown as solid lines.  Best-fit
parameters are provided in Table~\protect\ref{ta:eclipse}. 
\label{fig:eclipse} 
}
\end{figure}

Average eclipse light curves for each frequency were obtained by
combining results of every observed conjunction and are shown in
Figure~\ref{fig:eclipse}.  The light curve at 427~MHz has more scatter
than the other light curves, primarily due to higher levels of
sporadic, short-time scale radio interference of unknown origin,
particularly during the first orbit.

\begin{figure}
\plotone{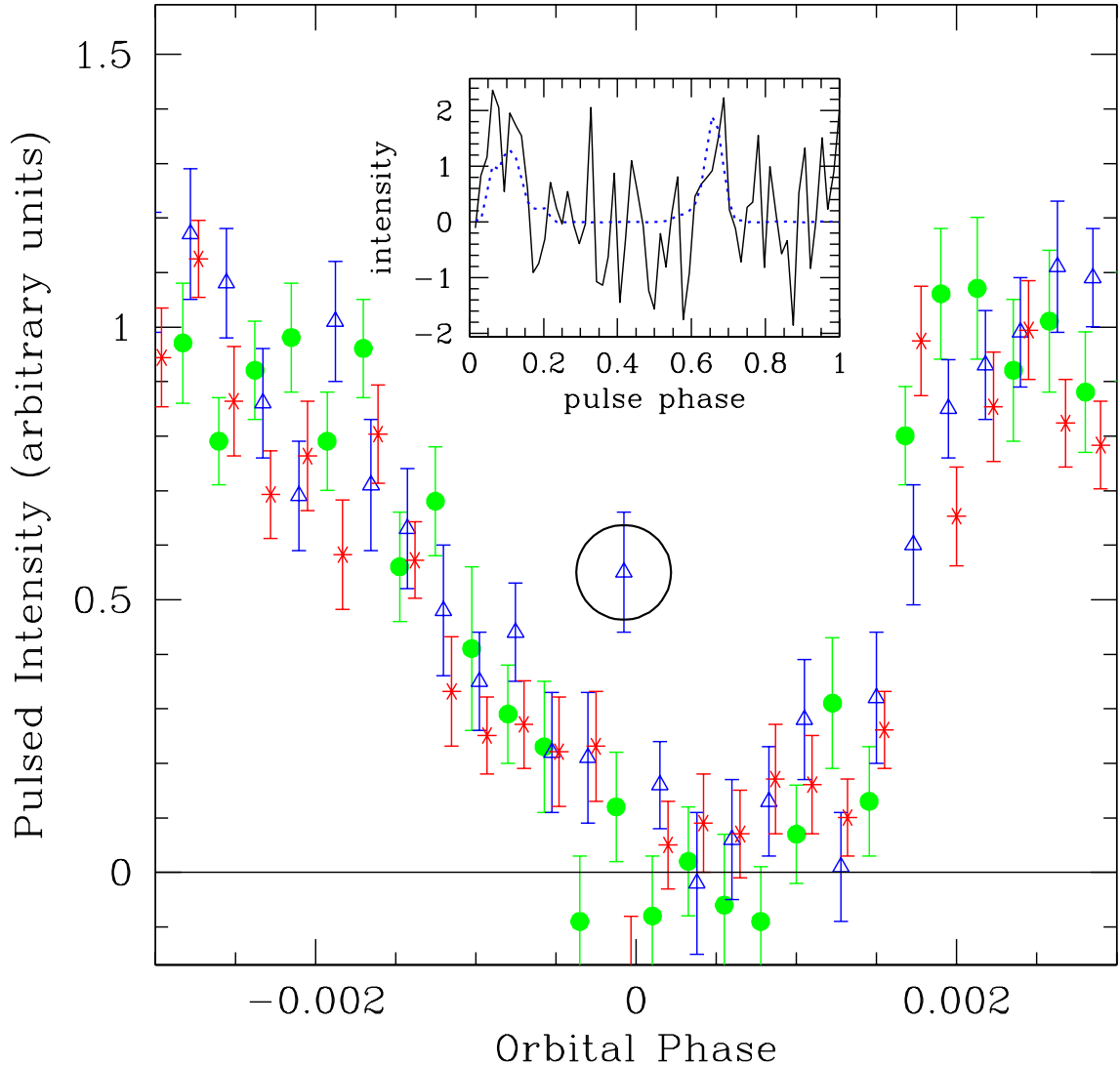}
\figcaption{
Close-up of eclipse region for the three orbits observed at 1400~MHz.
The first, second and third orbits are indicated with circles, triangles
and stars, respectively.  Note the circled second orbit point just before
conjunction; the pulsed intensity of this 2-s integration is $5\sigma$
above zero, indicating significant pulsed emission during eclipse.  The
pulse profile for this 2-s interval is shown in the inset (solid line),
as is the average template profile (dotted line), with the pre-determined
phase alignment.
\label{fig:variation}
}
\end{figure}

We detect significant variations in the pulsed emission at the same
orbital phase but in different orbits, during the eclipse.  An example
of such variability is illustrated in Figure~\ref{fig:variation}, in
which the light curves for the three 1400~MHz eclipses are shown.  In
the second observed orbit, there is an apparent $5\sigma$ detection of
the pulsar during the nominal eclipse period, as indicated in the
Figure.  Investigation of this 2-s interval's folded profile by eye
confirmed that the pulse, with the expected pulse shape and phase
(though appropriately noisy), was present (see
Fig.~\ref{fig:variation}, inset).  Also, in Figure~\ref{fig:eclipse},
there is emission in the eclipse region both before and after
conjunction in the average 427~MHz light curve.  Examination by eye
confirmed the approximate pulse shape at the expected phase in several
individual 2-s intervals in two of the three 427~MHz orbits.  No such
pulse appearance was seen in any of the 820~MHz eclipses, however.

Several new properties of the eclipse are apparent from Figure~1.  First, the 
eclipse is clearly asymmetric in that the eclipse ingress is longer in
duration than egress.  
Also, the eclipse appears to last longer post conjunction.

To quantify these properties, we fit the orbit-averaged light curves at
each frequency with two functions of the form
\begin{equation}
F(\phi) = \frac{1}{e^{(\phi - \phi_0)/w} + 1}.
\end{equation}
Here
$F(\phi)$ is the flux at orbital phase $\phi$, defined such that conjunction
is at $\phi = 0$.  This
function, having two free parameters $\phi_0$ and $w$, has value unity
for $(\phi - \phi_0)/w << -1$, zero for $(\phi - \phi_0)/w >> 1$, 
and one half for $\phi = \phi_0$.  
The transition from values near unity to zero occurs in
a width $\sim 2w$.  Optimal values of $\phi_0$ and $w$ for eclipse ingress
and egress were determined separately
for each light curve in Figure~1 using straightforward
$\chi^2$ minimization and data in the interval $-0.006 < \phi < 0$ for
ingress and $0 < \phi < 0.006$ for egress.  With two free parameters per
side, the number of degrees of freedom per fit, $\nu$, was 23.
The best fits are shown as solid curves in
Figure~1; they generally yield values of reduced $\chi^2$ near unity, 
except for the 427~MHz light curve, for which the higher $\chi^2$ results
from the extra scatter noted above.
Contours of constant $\chi^2$ in $\phi_0$/$w$ space
were approximately circular, indicating that these parameters
were largely independent.  Best-fit parameters and $\chi_{\nu}^2$ values are given in
Table~\ref{ta:eclipse}.

\begin{figure}
\plotone{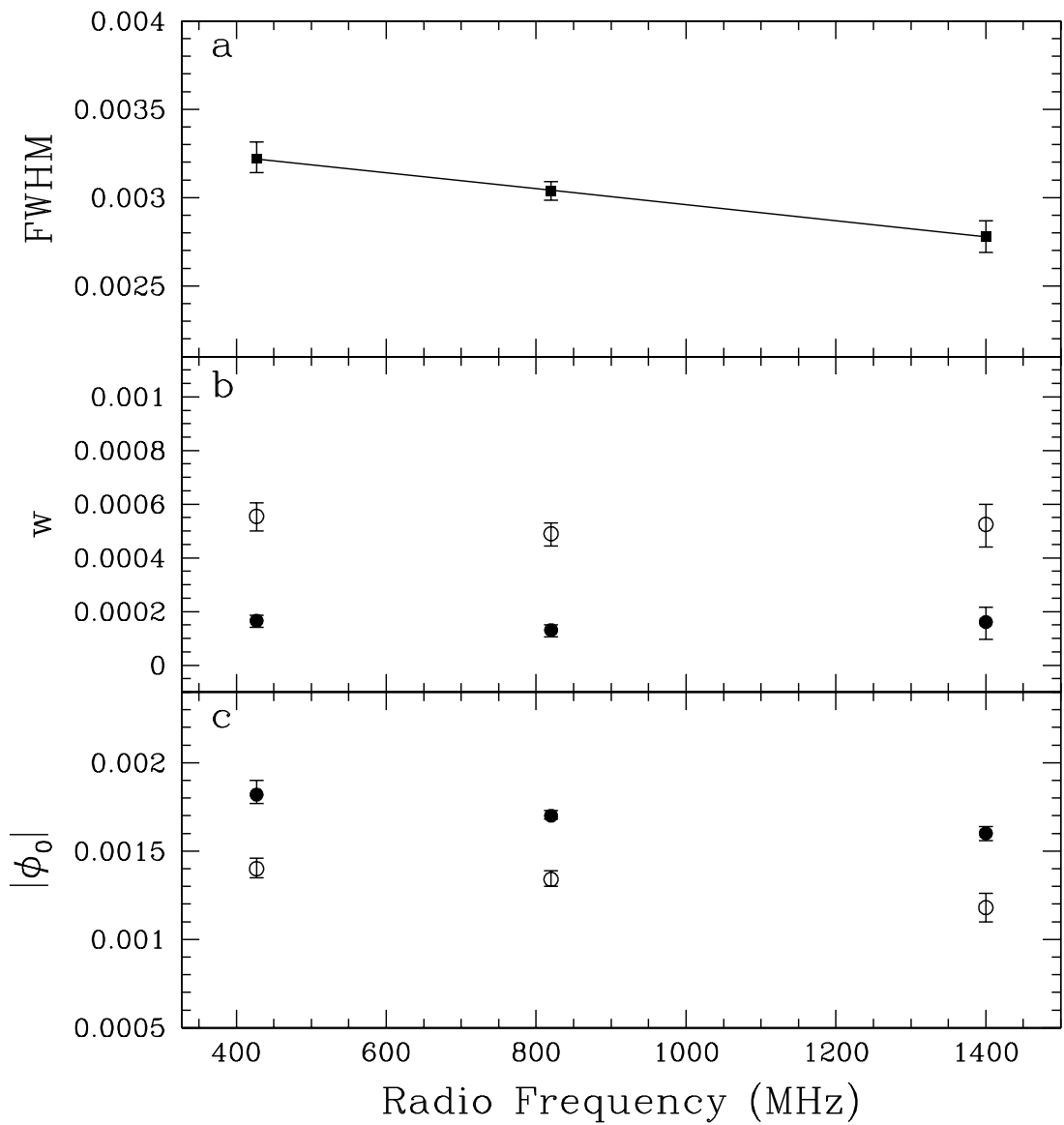}
\figcaption{Various fitted properties of pulsar A's eclipse as a function
of radio frequency.  Panel (a): eclipse duration (defined as the sum
of the ingress and egress values of $| \phi_0 |$, equivalent to FWHM), and best-fit linear
model (solid line).  Panel (b):
values of $w$ (approximately half the time for transition into and
out of eclipse)
for ingress (open symbols) and egress (filled symbols).
Panel (c): values of $| \phi_0 |$
for ingress (open symbols) and egress (filled symbols).
\label{fig:freq}
}
\end{figure}

Figure~\ref{fig:freq} summarizes the results of these fits.
Panel (a) shows the eclipse FWHM, defined as the sum of
ingress and egress values of $| \phi_0 |$.  The clear
trend in the data indicates that the eclipse duration is
frequency-dependent.  Indeed the reduced $\chi^2$ of a fit
to the data assuming a simple constant mean value is 6.1, 
clearly ruling this model out.  
The data are very well fit by a linear model,
having slope $(-4.52 \pm 0.03) \times 10^{-7}$~orbits~MHz$^{-1}$,
and intercept $(3.412 \pm 0.003) \times 10^{-3}$.
This predicts that the eclipse will disappear at approximately
7.5~GHz, assuming no other
eclipse mechanism becomes important.  
The mean eclipse FWHM duration, averaged over these three frequencies,
is $0.00301 \pm 0.00008$ in orbital phase, or $26.6 \pm 0.7$~s. 

Figure~\ref{fig:freq}, panel (b), shows ingress and egress values of $w$.
One aspect of the asymmetry of the eclipse is very apparent from this plot:
all ingress values of $w$ are higher than the egress values.  Thus,
the pulsar exits eclipse faster than it enters.  We quantify this effect
by taking the ratio of the ingress to egress values of $w$, $w_i/w_e$.
We find, for 427, 820 and 1400~MHz, $w_i/w_e = 3.34 \pm 0.59, 3.74 \pm 0.78,$ and
$3.45 \pm 1.40$, respectively.  Thus, this aspect of the eclipse asymmetry
is not dependent on frequency.  The mean $w_i/w_e$ is 3.51; the uncertainty
on this value, estimated from the uncertainties at each frequency, is 0.99.
Estimating the uncertainty from the RMS scatter in the three values suggests
a much smaller uncertainty of 0.15; this may be a chance occurence due to
small-number statistics.

Figure~\ref{fig:freq}, panel (c), shows the fitted values
of $\phi_0$ for ingress and egress at each frequency.
Clearly, $|\phi_0|$ values for egress are
systematically larger than for ingress (Fig.~2c),
demonstrating that the eclipse lasts longer post-conjunction.  
Note that the latter is {\it opposite} to what would be expected if the
$w$ asymmetry were the only one present, and hence is a different effect.
We quantify this using the difference between ingress and egress
values of $\phi_0$.  For 427, 820 and 1400~MHz, we find differences of
$0.00042 \pm 0.00007$, $0.00036 \pm 0.00006$, and $0.00042 \pm 0.00009$,
respectively.  Thus we find no evidence for frequency dependence of this
eclipse asymmetry.  That the absolute values of $\phi_0$ for both ingress
and egress decrease significantly with increasing frequency is a restatement of the fact
that the eclipse duration decreases with increasing frequency.

\section{Discussion}

The best-fit inclination angle of the system as reported by
\citet{rkr+04} is $88.7^{\circ} \pm 0.9^{\circ}$.  Given the measured
projected semi-major axis of pulsar A's orbit, this implies that A's
beam must pass within 0.07~lt-s of pulsar B.  The latter has
light-cylinder radius $R_{\rm lc} = cP/2\pi = 0.45$~lt-s.  Thus, if
B's magnetosphere were unperturbed, then A's beam as observed by us would have
passed well within its light cylinder and through its magnetosphere.
However, as discussed by \citet{lbk+04}, given the relative spin-down
luminosities of the two pulsars, with A's being a factor of $\sim$3625
greater than B's, it seems inevitable that A's wind has disrupted B's
outer magnetosphere at least halfway inside B's light cylinder and
possibly further out than A's beam passes at conjunction.
At A's conjunction, the relative transverse velocities of the pulsars is 
680~km~s$^{-1}$,
so the eclipse duration implies an approximately constant
physical size of the eclipsing region of 18,100~km, or 0.060~lt-s, comparable
to the estimated 0.07~lt-s impact parameter, though in a
direction parallel to the orbital plane.  This clearly implies an
eclipse region that is much smaller than B's unperturbed magnetosphere.

Synchrotron absorption in a shock heated ``magnetosheath'' surrounding
and containing B's magnetosphere is a possible model for the A eclipse
\citep[][Arons et al., in preparation]{asbk04a,lyu04}.  A's wind probably confines B's magnetosphere on the
side facing A (which changes periodically with B's rotation over 2.7~s)
such that the latter likely resembles a time-dependent variant of the
Earth's magnetosphere:  compressed on the side facing the source of the
wind (the Sun in the terrestrial case, A in this case) with a
magnetotail on the opposite side.  The plasma from A passes through a
bow shock which spreads and weakens with distance down the flanks of
the tail.  The wind plasma from A which passes through the bow shock,
flows in a magnetosheath layer between the bow shock and the
magnetopause current layer that confines B's magnetosphere.  If the bow
shock compresses and heats the A pulsar's wind plasma as if it occurred
in a weakly magnetized, semi-infinite flow, then synchrotron absorption
in this shock heated plasma can account for the strong eclipse
of A by B, albeit for surprisingly high pair densities in A's wind \citep{asbk04a}.
The dimensions of the magnetopause and bow shock on the
side facing A, along with the impact parameter of the line-of-sight
with respect to B at conjunction, are such that A's radio beam
can pass through the northernmost or southernmost extremities of the magnetosheath
layer (where north or south means with respect to the orbital plane.)

Such a scenario is consistent with the sharp edge and mildly
frequency-dependent shape of the eclipse, since the synchrotron
absorption optical depths at the observation frequencies used in this
study are large and the shock causes the plasma density and
relativistic temperature to jump up sharply across the thin shock
layer.  The eclipse-to-eclipse variations and occasional appearances of
the pulse during the nominal eclipse region suggest a changing eclipse
medium.  If pulsar B is an oblique rotator, it is possible that the
rotation of its magnetosphere causes the magnetospheric and bow shock
shapes to be asymmetric on average between ingress and egress by
amounts which account for the observed asymmetries.

If this picture and physical model is correct, one reaches the
interesting conclusion that A's wind shocks as if it is not strongly 
magnetized at a distance of only $\sim$750 times its light cylinder
radius.  Previous conclusions regarding pulsar wind magnetization have
been derived from the behavior of relativistic winds as they pass
through their termination shocks in pulsar wind nebulae \citep[see][for
a recent review]{krh04}, located at much larger distances (typically
$\sim 10^7$ light cylinder radii) from their driving pulsars
\citep[e.g.][]{kc84a}.

The above A eclipse model (see Arons et al., in preparation) predicts that 
at sufficiently
high radio frequencies, the eclipse will clear, consistent with the
observed frequency dependence of the eclipse duration.  For nominal,
highly simplified models, transparency might appear at frequencies
above 5--10~GHz, roughly quantitatively consistent with the observed
frequency dependence extrapolation.  If the impact parameter of the
line-of-sight with respect to B during eclipse is sufficiently small,
cyclotron scattering within B's magnetosphere might also contribute to
the high-frequency opacity.  However, nominal values for the geometry
suggest this to be unlikely.  In addition, the eclipse duration and
morphology should change with time on few-year time scales as
relativistic periastron precession, and, perhaps more importantly,
geodetic precession, which is expected to be very important in this
system \citep{lbk+04}, change the alignment of the pulsars.  This will
alter both our viewing vantage point and the dynamic pressure of A's
wind on B at conjunction.

The durations of pulsar A ingress and egress are comparable to the
rotational period of B.  As B's magnetosphere could well present a
different response to A's wind pressure depending on B's rotational
phase, significantly different A eclipse behavior from orbit to orbit
might be expected.  A higher time resolution investigation with
higher sensitivity may reveal B phase dependent effects on the eclipse.

We pay tribute to the Parkes survey team for so wonderful a discovery.
We thank C. Bignell, F. Ghigo, and G. Langston for assistance with the
GBT observations, and M. Roberts and B. Rutledge for useful discussions.
This work was funded by an NSERC Discovery Grant and Steacie Fellowship
Supplement to V.M.K.  Additional funding to V.M.K. was provided by
FQRST and CIAR.  J.A. was partially supported by NASA grants TM4-5005X,
NAG5-12031, G03-4063B and HST-HF-01157.01-A. He also thanks the taxpayers
of California for their continued indulgence.  A.S. acknowledges support
provided by NASA through Chandra Fellowship PF2-30025
awarded by the Chandra X-ray Center, which is operated by the Smithsonian
Astrophysical Observatory for NASA under contract NAS8-39073.

{\it Note added in proof. -- } Very recently, McLaughlin et al. (2004)
report modulation of A's flux at B's period during eclipse ingress, from
a higher time resolution analysis of the 820 MHz data set used here.
This suggests that the eclipse medium is indeed highly dependent on B's
rotational phase.


\begin{thebibliography}{8}
\expandafter\ifx\csname natexlab\endcsname\relax\def\natexlab#1{#1}\fi

\bibitem[{Arons {et~al.}(2004a)Arons, Spitkovsky, Backer, \& Kaspi}]{asbk04a}
Arons, J., Spitkovsky, A., Backer, D.~C., Kaspi, V. M. 2004, in
Proc. 2004 Apsen Winter Conf. on Astrophysics, ``Binary Radio Pulsars,''
eds. F. Rasio \& I. Stairs, PASP, in press (astro-ph/0404159)

\bibitem[{Backer {et~al.}(1997)Backer, Dexter, Zepka, D., Wertheimer, Ray, \&
  Foster}]{bdz+97}
Backer, D.~C., Dexter, M.~R., Zepka, A., D., N., Wertheimer, D.~J., Ray, P.~S.,
  \& Foster, R.~S. 1997, PASP, 109, 61

\bibitem[{{Burgay} {et~al.}(2003){Burgay}, {D'Amico}, {Possenti}, {Manchester},
  {Lyne}, {Joshi}, {McLaughlin}, {Kramer}, {Sarkissian}, {Camilo}, {Kalogera},
  {Kim}, \& {Lorimer}}]{bdp+03}
{Burgay}, M., {D'Amico}, N., {Possenti}, A., {Manchester}, R.~N., {Lyne},
  A.~G., {Joshi}, B.~C., {McLaughlin}, M.~A., {Kramer}, M., {Sarkissian},
  J.~M., {Camilo}, F., {Kalogera}, V., {Kim}, C., \& {Lorimer}, D.~R. 2003,
  Nature, 426, 531

\bibitem[{Kaspi {et~al.}(2004)Kaspi, Roberts, \& Harding}]{krh04}
Kaspi, V.~M., Roberts, M. S.~E., \& Harding, A.~K. 2004, in Compact Stellar
  X-ray Sources, ed. W.~H.~G. Lewin \& M.~van~der Klis (United Kingdom:
  Cambridge University Press), in press

\bibitem[{Kennel \& Coroniti(1984)}]{kc84a}
Kennel, C.~F. \& Coroniti, F.~V. 1984, ApJ, 283, 694

\bibitem[{Lyne {et~al.}(2004)Lyne, Burgay, Kramer, Possenti, Manchester,
  Camilo, McLaughlin, Lorimer, D'Amico, Joshi, Reynolds, \& Freire}]{lbk+04}
Lyne, A.~G., Burgay, M., Kramer, M., Possenti, A., Manchester, R.~N., Camilo,
  F., McLaughlin, M.~A., Lorimer, D.~R., D'Amico, N., Joshi, B.~C., Reynolds,
  J.~R., \& Freire, P. C.~C. 2004, Science, 303, 1153

\bibitem[{Lyutikov(2004)}]{lyu04}
Lyutikov, M. 2004, MNRAS, submitted (astro-ph/0403076)

\bibitem[{McLaughlin {et~al.}(2004)}]{m++04}
McLaughlin, M. A., et al. 2004, ApJ, submitted (astro-ph/0408297)

\bibitem[{Ransom(2001)}]{ran01}
Ransom, S.~M. 2001, PhD thesis, Harvard University

\bibitem[{Ransom {et~al.}(2004)Ransom, Kaspi, Ramachandran, Demorest, Backer, Pfahl, Ghigo, \& Kaplan}]{rkr+04}
Ransom, S. M., Kaspi, V. M., Ramachandran, R., Demorest, P., Backer, D. C., Pfahl, E. D., Ghigo, F. D., \& Kaplan, D. L. 2004, ApJ, 609, L71

\end{thebibliography}

\newpage

\begin{table}[t]
\begin{center}
\caption{Summary of GBT Observations of \psr}
\begin{tabular}{ccccccc} \hline\hline
Date & Start UTC & Start MJD & Duration & Frequency & Bandwidth & No. Channels\\
(dd/mm/yy) & & & (hr) & (MHz) & (MHz) & \\\hline
11/12/03 & 04:36:27  & 52984.19199 & 6.15 & 1400 & 96 & 96 \\
18/12/03 & 04:12:15  & 52992.17519 & 5.30 & 820  & 48 & 96 \\
23/12/03 & 05:22:22  & 52997.22388 & 4.72 & 820  & 48 & 96 \\
01/01/04 & 03:30:09  & 53005.14595 & 6.22 & 427  & 48 & 96 \\\hline
\end{tabular}
\label{ta:obs}
\end{center}
\end{table}

\begin{table}[t]
\begin{center}
\caption{Fit Parameters for the \psr A Eclipse}
\begin{tabular}{ccccc|ccc} \hline\hline
Frequency & No. Eclipses & \multicolumn{3}{c}{Ingress} & \multicolumn{3}{c}{Egress} \\ \cline{3-5}\cline{6-8}
 & Observed & $\phi_0$ & $w$ & $\chi^2_{23}$ &  $\phi_0$ & $w$ & $\chi^2_{23}$ \\
(MHz)  &  & ($\times 10^{-3}$) & ($\times 10^{-3}$) &  & ($\times 10^{-3}$) & ($\times 10^{-3}$) & \\\hline  
427 & 3 & $-1.60^{+0.05}_{-0.06}$ & $0.56^{+0.06}_{-0.05}$ & 2.18 & $1.82^{+0.03}_{-0.08}$ & $0.17^{+0.03}_{-0.02}$ & 1.65   \\
820 & 4 & $-1.34^{+0.04}_{-0.05}$ & $0.49^{+0.05}_{-0.04}$ & 0.834 & $1.70^{+0.02}_{-0.03}$ & $0.13^{+0.03}_{-0.02}$ & 0.76 \\
1400 & 3 & $-1.18^{+0.08}_{-0.06}$ & $0.53^{+0.08}_{-0.08}$ & 0.53 & $1.60^{+0.04}_{-0.04}$ & $0.16^{+0.07}_{-0.06}$ & 1.02 \\ \hline
\multicolumn{8}{c}{Note: $\chi^2_{23}$ is the reduced $\chi^2$ for 23 degrees of freedom.}
\end{tabular}
\label{ta:eclipse}
\end{center}
\end{table}

\clearpage

\clearpage

\clearpage

\end{document}